\documentclass[doublecol]{epl2} 
\usepackage{enumerate}
\usepackage{amsmath}
\usepackage{amsfonts}
\usepackage{amssymb}
\usepackage[utf8]{inputenc}
\usepackage[T1]{fontenc}
\usepackage{mathtools}
\usepackage{wasysym}
\usepackage{accents}
\usepackage[english]{babel}
\usepackage[svgnames]{xcolor}
\usepackage[colorlinks=true,citecolor=blue]{hyperref}
\usepackage{url}
\usepackage{float}
\RequirePackage{graphicx}
\usepackage{amsmath}
\usepackage[version=4]{mhchem} 

\newcommand{\p}{p_{\text{P}}}
\newcommand{\vg}{V_{\text{gr}}}

\title{Fractional quantum mechanics meets quantum gravity phenomenology}
\shorttitle{Fractional quantum mechanics meets quantum gravity phenomenology} 

\author{Gislaine Varão\inst{1} \and Iarley P. Lobo\inst{2} \and Valdir B. Bezerra\inst{1}}
\shortauthor{Gislaine Varão \etal}

\institute{                    
  \inst{1} Physics Department, Federal University of Para\'iba - Caixa Postal 5008, 58059-900, Jo\~ao Pessoa, PB, Brazil.\\
  \inst{2} Department of Chemistry and Physics, Federal University of Para\'iba - Rodovia BR 079 - km 12, 58397-000 Areia-PB,  Brazil.
}

\abstract{
This letter extends previous findings on the modified Schrödinger evolution inspired by quantum gravity phenomenology. By establishing a connection between this approach and fractional quantum mechanics, we provide insights into a potential deep infrared regime of quantum gravity, characterized by the emergence of fractal dimensions, similar to behaviors observed in the deep ultraviolet regime. Additionally, we explore the experimental investigations of this regime using Bose-Einstein condensates. Notably, our analysis reveals a direct implication of this analogy: general experiments probing fractional quantum mechanics may serve as equivalent models of quantum gravity. We identify instances of nonlocal behavior in such systems, suggesting an analogous phenomenon of nonlocality in quantum gravity.
}

\begin{document}

\maketitle

\section{Introduction}
\label{intro}

The quantization of spacetime has been a formidable challenge over the past several decades. Various approaches have been developed in the quest for a theory that accurately describes gravitational degrees of freedom \cite{Kiefer:2004xyv,Polchinski:1998rq,Ashtekar:2021kfp,Ashtekar:2021kfp,Loll:2019rdj,Eichhorn:2018yfc}, among others. These developments have paved the way for phenomenological approaches to quantum gravity, which have matured to a point where they can now be explored through experimental and observational means. These approaches typically involve modeling common properties with free parameters constrained by experimental data or observations (see \cite{Amelino-Camelia:2008aez,Addazi:2021xuf,Mattingly:2005re} for comprehensive reviews).

Astrophysics has been a particularly fruitful arena for probing the quantum nature of spacetime and its symmetries due to the high energies achievable by astroparticles \cite{Amelino-Camelia:2008aez,Mattingly:2005re}. However, this field is often plagued by significant uncertainties and  model-dependence analyses \cite{AlvesBatista:2023wqm}. Conversely, tabletop experiments, conducted in controlled environments, allow for the measurement of observables with much higher precision despite the lower energies involved.

Astrophysical observations are effective at detecting deviations from standard physics that scale with high powers of momentum (testing the ultraviolet (UV) regime), but they often miss corrections that scale with lower powers, such as linear corrections. In this context, an intriguing possibility is to explore quantum gravity in the infrared (IR) regime, where it manifests as deformed quantum mechanics \cite{Amelino-Camelia:2008aez}. Recent studies have demonstrated that such IR departures can provide bounds competitive with those from high-energy physics, achieving Planck-scale sensitivity \cite{Amelino-Camelia:2008aez,Wagner:2023fmb}. In fact, the analysis of fluctuations of spacetime geometry due to quantum gravity effects suggest a quantum spacetime with fractional powers in the Planck scale parameter \cite{Sorkin:1996sr,Marolf:2003bb,Keshet:2024hlm,Yunes:2016jcc}.

Therefore, we pose the following question. What happens if the corrections in the Schr\"{o}dinger evolution due to modified dispersion relations involve noninteger powers? The replacement of the Laplacian in the Schrödinger equation with a version featuring half-integer powers of the momentum operator, known as fractional quantum mechanics \cite{Laskin:1999tf,laskin2000fractional,laskin2018fractional}, introduces a fractal structure to the dimensions of the space under investigation. This raises the question: what impact would such inclusions have if they were connected to quantum gravity-inspired models? Could the fundamental quantum spacetime possess a fractal structure? And what about its nonrelativistic limit? These are the questions we aim to explore in this letter.


\section{Modified dispersion relations in the nonrelativistic limit}\label{sec:mdr}

The most studied aspect of quantum gravity phenomenology involves the analysis of deviations from the relativistic dispersion relation due to Planck-scale-suppressed contributions. These contributions describe the interaction of test particles with quantized degrees of freedom of gravity, characterized by the Planck energy $E_{\text{P}} = \sqrt{\hbar c^5/G} \approx 1.2 \times 10^{19}$ GeV or the Planck momentum $\p = E_{\text{P}}/c \approx 6.5$ N$\cdot$s, where $c$ is the speed of light, $G$ is the gravitational constant, and $\hbar$ is the reduced Planck constant.

In this letter, we focus on the simplest version of Modified Dispersion Relation (MDR) with corrections linear in $\p^{-1}$ that include powers of the energy, momentum, and mass of the test particle \cite{Wagner:2023fmb}:
\begin{align} \label{eq:mdr1}
&M^2 c^2 = \left( \frac{E}{c} \right)^2 - p^2\nonumber \\
&- \p^{-1} \sum_{n=0}^{2} \sum_{m=0}^{3-n} a_{n,m,3-n-m} (Mc)^n p^m \left( \frac{E}{c} \right)^{3-n-m},
\end{align}
where $M$ is the mass of the particle, $E$ is its energy, and $p$ is its momentum. This is a general form of MDR, that involves a wide range of cases that are relevant for both high or low energy probes.

These modifications not only align with the expected interaction with quantum gravitational degrees of freedom, but they also emerge from the semiclassical limit of various quantum gravity theories \cite{Lukierski:1992dt,Majid:1994cy,Amelino-Camelia:2016gfx,Amelino-Camelia:1996bln,Freidel:2005me}. Experimentally, higher-order contributions in $E_{\text{P}}^{-1}$ are often negligible, as most tests do not reach Planck-scale sensitivity even for $E_{\text{P}}^{-2}$ corrections \cite{Amelino-Camelia:2008aez,Addazi:2021xuf,AlvesBatista:2023wqm}.

The MDR in \eqref{eq:mdr1} encompasses integer powers of $M$, $p$, and $E$. For nonrelativistic analyses, the primary contributions are typically linear or quadratic in momentum. A generic form of the dispersion relation in the nonrelativistic regime is given by \cite{Wagner:2023fmb}:
\begin{equation}\label{eq:mdr2}
E = \frac{p^2}{2M} + \frac{\p^{-1}}{2M} \sum_{n=1}^{3} \xi_n (Mc)^{3-n} p^n,
\end{equation}
where $\xi_n$ is a linear combination of $a_{n,m,3-n-m}$.

We have demonstrated that the quantization of this expression resultes in a modified Schrödinger equation that includes corrections involving both integer and half-integer powers of the Laplacian operator $\Delta$ \cite{Wagner:2023fmb}.

One of the key results from \cite{Wagner:2023fmb} is that, for the harmonic oscillator, it is possible to place rough constraints on the parameters $\xi_n$ of the MDR \eqref{eq:mdr2}: $\xi_1 \leq \mathcal{O}(10^{-1})$, $\xi_2 \leq \mathcal{O}(10^{10})$, and $\xi_3 \approx \mathcal{O}(10^{21})$. These bounds improve upon previous estimates such as $\xi_1 \leq \mathcal{O}(1)$ obtained using cold atoms \cite{Amelino-Camelia:2009wvc}, which had already shown the capability of these tests in reaching Planck scale sensitivity.


\section{Fractional Powers of Planck-Scale Corrections from Fractal Black Holes}\label{sec:barrow}

Many approaches to quantum gravity predict corrections that grow with integer powers of the Planck scale. However, it is intriguing that there are scenarios in which corrections proportional to fractional powers of the Planck scale are present due to quantum gravity effects. For instance, fluctuations in the horizon radius of quantum black holes could be of the order $\ell_{\text{P}}^{2/3}$ \cite{Sorkin:1996sr} or $\sqrt{\ell_{\text{P}}}$ \cite{Marolf:2003bb}, where $\ell_{\text{P}} = \sqrt{\hbar G / c^3} \sim 10^{-35}$ m is the Planck length. These fluctuations might even result in Hawking radiation, whose experimental verification is discussed in \cite{Keshet:2024hlm}. Such fractional powers indicate a fractal nature of the quantum black hole horizon, suggesting that quantum gravity effects increase the effective dimension of the horizon beyond the conventional 2-dimensional surface.

Recently, Barrow proposed that the fractal behavior of black hole horizons and their thermodynamics could be related to the fractional nature of gravity. In \cite{Barrow:2020tzx} it is suggested an entropy of the form $S_{\text{frac}} = (A / 4\ell_{\text{P}})^{d/2}$, where $A = 4\pi r_h^2$ is the black hole area with horizon radius $r_h$, and $d \geq 2$ represents the fractal dimension of the black hole horizon.

The relationship between modified black hole thermodynamics and modified dispersion relations (or generalized uncertainty principles) has been known for some time \cite{Adler:2001vs,Amelino-Camelia:2005zpp,Ling:2005bq}, and we can find which MDR could describe this fractional law. To estimate Barrow's black hole entropy, we consider a fractional modified dispersion relation for photons of the form
\begin{equation}\label{eq:barrow-energy}
    E = \frac{2^{d-1}}{d} \left( \frac{G}{\pi} \right)^{\frac{d}{2}-1} p^{d-1} \equiv \xi_d \ell_{\text{P}}^{d-2} p^{d-1},
\end{equation}
where the dimensionless terms are absorbed into $\xi_d$ and $\ell_{\text{P}} = \sqrt{G}$ (with $c = \hbar = 1$), where $d$ will play the role of the fractal dimension.

Due to Hawking radiation, there is a characteristic uncertainty in the position of a particle of the order of the Schwarzschild radius, $\Delta x = r_h = 2GM$. Comparing this with the Heisenberg uncertainty principle leads to an uncertainty in the particle's momentum of $\Delta p \approx 1 / (4GM)$. Relating the momentum scale $\Delta p \sim p$ to an energy scale $E$ via \eqref{eq:barrow-energy}, and recognizing that this energy is proportional to the Hawking temperature $T$, we obtain
\begin{equation}
    T = \frac{1}{d(4\pi G)^{d/2}} \frac{1}{M^{d-1}}.
\end{equation}

Using the first law of black hole thermodynamics $dM = TdS$, we integrate to find the entropy
\begin{equation}
    S = (4\pi G)^{d/2} M^d = \left( \frac{A}{4\ell_{\text{P}}} \right)^{d/2},
\end{equation}
where $A=4\pi r_h^2$. This shows a close link between modified dispersion relations with fractional powers of momentum and the fractional structure of the black hole horizon, aligning with quantum gravity effects.

Although such scenarios are not usually considered in most astrophysical analyses, the gravitational wave community has shown interest, as noted in \cite{Yunes:2016jcc} (equation (22)). This has been theoretically investigated under ``Multifractional Spacetime Theory'' \cite{Calcagni:2009kc}. 
Since gravitational waves are typically detected in a low-energy regime, fractional powers in the dispersion relation can be tested, with less than quadratic powers potentially competing with standard quadratic terms.

To explore a broader range of phenomenological possibilities, we propose examining the implications of assuming non-integer values of momentum powers in the nonrelativistic dispersion relation. While the impact of non-integer powers on astrophysical tests is straightforward, their inclusion in quantum mechanical modifications of the Galilean dispersion relation is more complex, as it implies fractional powers of the Laplacian. This observation, which applies when $n$ is an odd integer, was not extensively explored in \cite{Wagner:2023fmb}. We aim to extend this analysis to non-integer values, considering the first quantization of a modified dispersion relation (MDR). To phenomenologically incorporate fractional powers of the Planck scale, we consider the following MDR:
\begin{equation}\label{eq:mdrgen0}
    E^2 = c^2 p^2 + M^2 c^4 + \xi_{\alpha,\beta} \ell_{\text{P}}^{-\beta} c^2 (Mc)^{2+\beta-\alpha} p^{\alpha}.
\end{equation}

Here, $\beta$ is a real number and regulates the fractional power of the Planckian units that appears in the MDR. As we discussed in this section, quantum gravity corrections with this property are a physical possibility, therefore we add $\beta$ as phenomenological parameter that should be experimentally constrained. Intuitively, it is connected to the possibility of a fractal nature of event horizons. If $\beta$ is an integer, the MDR might originate from a Taylor expansion involving powers of $p \ell_{\text{P}}$. If $\beta$ is a half-integer, the MDR may arise from a Puiseux series, a generalization of the Taylor expansion with fractional powers. The nonrelativistic expansion of this MDR gives a correction to the kinetic energy as
\begin{equation}\label{eq:mdrgen}
    E = \frac{p^2}{2M} + \xi_{\alpha,\beta} \ell_{\text{P}}^{-\beta} \frac{(Mc)^{2+\beta-\alpha}}{2M} p^{\alpha} \equiv \frac{p^2}{2M} + D_{\alpha}^{(\beta)} p^{\alpha}.
\end{equation}

In this expression, $\alpha$ is a parameter that can take positive real values. Notice that if $\beta = 1$, and $\alpha = \{1, 2, 3\}$, we recover the corrections in \eqref{eq:mdr2}. Our correction factors will be captured in the definition of $D_{\alpha}^{(\beta)}$ in \eqref{eq:mdrgen}.


\section{Fractional quantum mechanics and quantum gravity phenomenology}\label{sec:frac}

We rely on the historical notes presented in the preface of \cite{laskin2018fractional}. Shortly after the discovery of path integral quantization, Mark Kac demonstrated that performing a Wick rotation in time transforms Feynman's path integral into Wiener's path integral. Since the latter is an integral over Brownian motion, Feynman's approach is often referred to as Brownian-like path quantization. In Brownian motion, the position of diffusive particles evolves with the square root of time, $x(t) \sim t^{1/2}$. However, this is not the most general form of diffusive behavior. For many complex phenomena, the position of diffusive particles follows a scaling law $x(t) \sim t^{1/\alpha}$, where $0 < \alpha \leq 2$. The model describing this type of scaling is known as Lévy flights, with $\alpha$ referred to as the Lévy index. When $\alpha = 2$, one recovers Brownian motion.

In \cite{Laskin:1999tf}, Laskin generalized Feynman's path integral quantization to include Lévy flights. This generalization resulted in a modified Schrödinger equation that could also be derived from the first quantization of the Hamiltonian $H = D_{\alpha}p^{\alpha} + V$, where $D_{\alpha}$ is a parameter and $V$ is the potential. The corresponding pure fractional Schrödinger equation is
\begin{equation}\label{eq:pure_frac_sch1}
    i\hbar\frac{\partial}{\partial t}\Psi({\bf x},t) = \left( D_{\alpha}(-\hbar^2 \Delta)^{\alpha/2} + V({\bf x},t) \right) \Psi({\bf x},t).
\end{equation}

The term appearing in the free part of Eq.\eqref{eq:pure_frac_sch1} is derived from the quantization of Eq.\eqref{eq:mdr2} and includes a potential term:
\begin{align}\label{eq:main_schr}
i\hbar\frac{\partial}{\partial t}\Psi({\bf x},t) &= \left(-\frac{\hbar^2}{2M}\Delta + V({\bf x},t)\right. \\
&\left.+\xi_{\alpha,\beta}\p^{-\beta}\frac{(Mc)^{2-\alpha+\beta}}{2M} \left( -\hbar^2\Delta \right)^{\alpha/2}\right)\Psi({\bf x},t)\, .\nonumber
\end{align}

Despite the limited range for the parameter $\alpha$ ($0 < \alpha \leq 2$), fractional techniques are well-suited for our purposes. Corrections for $n > 2$ in \eqref{eq:mdr2} are far from current Planck-scale sensitivity, as shown in \cite{Wagner:2023fmb}. Therefore, we will consider this parameter interval.

We can interpret fractional differential operators using the Riesz derivative \cite{laskin2000fractional}, defined as: 
\begin{equation}
    \Delta^{\alpha/2}\Psi(\vec{x},t) = -\frac{1}{(2\pi)^3}\int d^3k\, e^{i\vec{k}\cdot \vec{x}} k^{\alpha} \Phi(\vec{p},t),
\end{equation}
where the functions $\Psi(\vec{x},t)$ and $\Phi(\vec{p},t)$ are related by Fourier transforms:
\begin{equation}
    \Phi(\vec{p},t) = \int d^3x\, e^{-i \vec{p} \cdot \vec{r}} \Psi(\vec{x},t).
\end{equation}
Using the definition of the Riesz derivative, it has been shown that the fractional part of the deformed Hamiltonian is Hermitian and preserves parity \cite{laskin2018fractional}. Consequently, the new Hamiltonian operator retains these properties. Since this is a nonlocal operator, we should expect that delocalization of particles will be present in such spacetime. This expectation will be confirmed by experimental analogues that we will discuss in the last section.


\subsection{The deep infrared regime}

Many phenomenological analyses of Lorentz Invariance Violation (LIV) do not focus on the specific theory from which a Modified Dispersion Relation is derived. The primary objective is to detect any signs of anomalous results that deviate from those predicted by special relativity. Therefore, it is reasonable to assume an MDR of the form $E^2 - c^2p^2 = c^4M^2 + \xi^n (cp)^{n+2} / E_{\text{P}}^n$ as an exact expression, to be directly applied throughout an analysis \cite{Amelino-Camelia:2013tla}.

Additionally, similar studies are conducted from a theoretical perspective on the effects of dimensional reduction in quantum gravity. In this context, exact dispersion relations are used to study diffusion processes \cite{Benedetti:2008gu,Horava:2009if,Amelino-Camelia:2013tla,Calcagni:2013vsa,Sotiriou:2011aa} or from the standpoint of thermodynamics \cite{Amelino-Camelia:2016sru,Lobo:2020oqb}. These studies have shown that the spacetime dimension fluctuates, imparting a fractal nature to the geometry, reaching asymptotically certain values in the deep UV or trans-Planckian regime.

In our study, we considered a relativistic MDR of the form \eqref{eq:mdr1} and derived a nonrelativistic Galilean Invariance Violating MDR \cite{Bosso:2023nst}, from which first quantization was performed. From a pragmatic phenomenological perspective, one might reason that the MDR \eqref{eq:mdrgen} could be assumed as an exact expression for elementary particles, albeit suppressed by fractional powers of the Planck scale. For a composite system composed of $N$ elementary particles, there are several arguments supporting similar correction, but suppressed by a scale of $N\p$ \cite{Amelino-Camelia:2013fxa}.


A natural consequence of this assumption is that a MDR of the form \eqref{eq:mdrgen} is dominated by the term involving $p^{\alpha}$ for sufficiently small spatial momenta, when $\alpha < 2$. We refer to the regime where the Planckian term in \eqref{eq:mdr2} becomes dominant as the \textit{deep infrared regime}.

In terms of the momentum of a particle of mass $M$, from \eqref{eq:mdr2}, this regime is characterized by the following inequality:
\begin{equation}\label{eq:ineq1}
    p \ll \left[ 2MD^{(\beta)}_{\alpha} \right]^{1/(2-\alpha)}.
\end{equation}

Thus, the MDR in this regime simplifies to the pure fractional expression:
\begin{equation}\label{eq:mdr3}
    E = D^{(\beta)}_{\alpha} p^{\alpha}=\xi_{\alpha,\beta}\p^{-\beta}\frac{(Mc)^{2+\beta-\alpha}}{2M}p^{\alpha}\, .
\end{equation}

In this regime, we are dealing with pure fractional quantum mechanics, and we can directly apply the results from this scenario to quantum gravity phenomenology. 
Consider a cooled system of quantum particles with $N$ constituents with mass $M$, such as a Bose-Einstein condensate. In this case, the Planck momentum scales as $p_{\text{P}} \rightarrow Np_{\text{P}}$. From fractional statistical mechanics \cite{laskin2018fractional}, the internal energy of an ideal gas is given by $U = 3N k_{\text{B}}T / \alpha$, where $k_{\text{B}}$ is Boltzmann's constant. Assuming the fractional Bose gas has a similar scaling equation of state, the energy of this system is approximately:
\begin{equation}\label{eq:energy_temp}
    E \sim N \frac{k_{\text{B}}T}{2\alpha}.
\end{equation}

Combining \eqref{eq:ineq1}, \eqref{eq:mdr3}, and \eqref{eq:energy_temp}, we find that the resulting expression is independent of the number of particles, leading to the following condition:
\begin{equation}\label{eq:temp1}
    T \ll T_{\text{dir}} = \frac{\alpha}{Mk_{\text{B}}} \left[ \xi_{\alpha,\beta} (Mc)^{2+\beta-\alpha}\p^{-\beta} \right]^{\frac{3-\alpha}{2-\alpha}},
\end{equation}
where $T_{\text{dir}}$ represents the upper temperature limit for the deep infrared regime. This implies that a system with temperature $T \ll T_{\text{dir}}$ should be governed by fractional quantum mechanics in this scenario. Therefore, if one prepares a system at very low temperatures and observes no deviations from standard quantum mechanics, it suggests that the fractional temperature ceiling $T_{\text{dir}}$ must be at least on the same order, or at most, one or two orders of magnitude larger than the achieved temperature. This allows us to estimate the experimentally allowed region in the parameter space of $(\alpha,\beta)$. 

To illustrate this discussion, we refer to the results of \cite{Deppner:2021fks}, which achieved a record low temperature of approximately $10^{-12}$ K using a Bose-Einstein condensate. The experimental setup involves a Bose-Einstein condensate composed of $10^5$ atoms of \ce{^{87}Rb}, with the mass of \ce{^{87}Rb} being $M = 1.4 \times 10^{-25}$ kg. Additionally, we use the speed of light $c = 3 \times 10^8$ m/s, Boltzmann's constant $k_{\text{B}} = 1.4 \times 10^{-23}$ J/K, the Planck momentum $p_{\text{P}} = 6.5$ N$\cdot$s, and assume that each atom contains approximately 100 elementary particles, resulting in an effective Planck momentum of $650$ N$\cdot$s, with $\xi_{\alpha,\beta} = 1$.

\begin{figure}[h]
    \centering
    \includegraphics[width=\columnwidth]{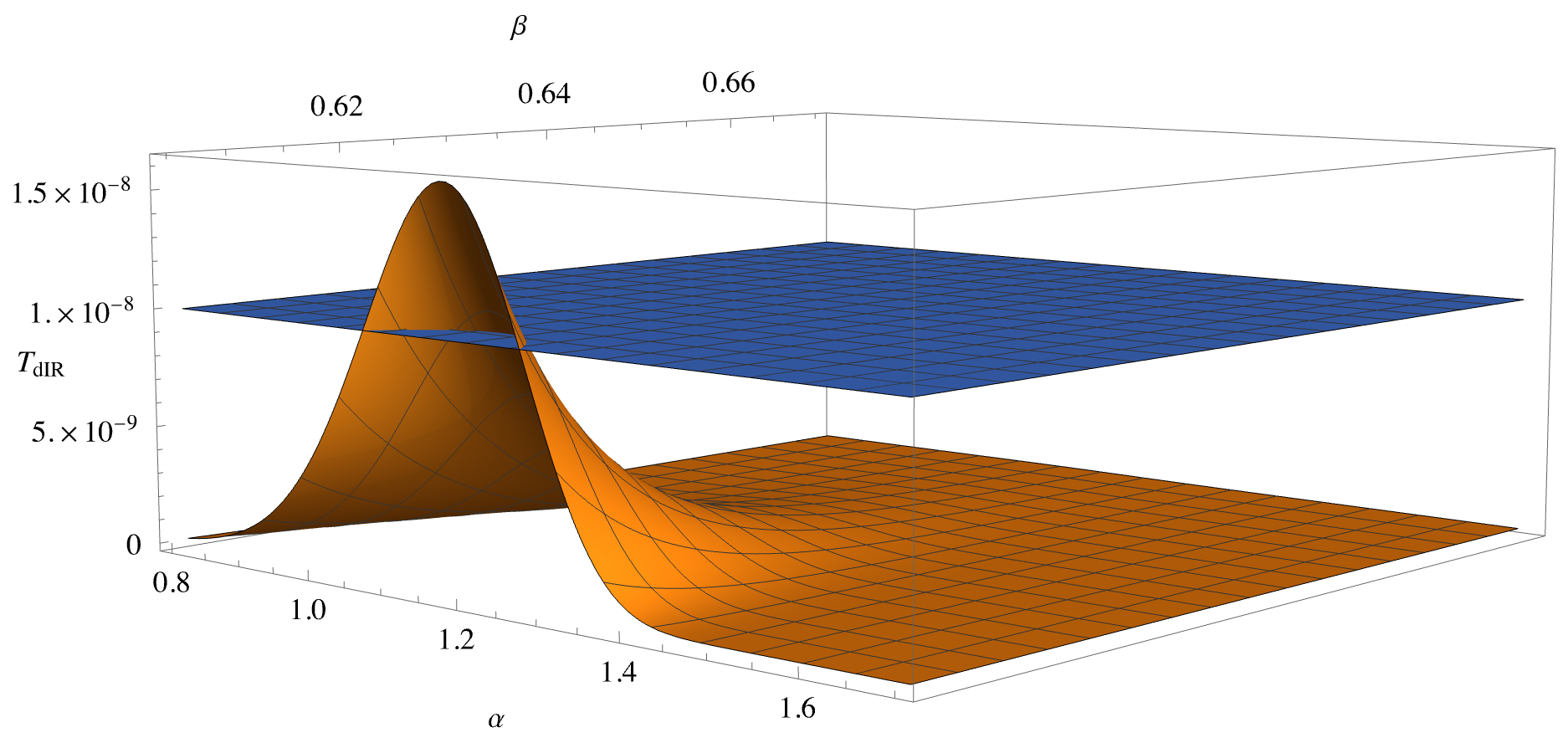}
    \caption{The dependence of $T_{\text{dir}}$ on the parameters $\alpha$ and $\beta$ is depicted by the orange surface, while the safe ceiling is indicated by the blue plane.}
    \label{fig:temperature-dir}
\end{figure}

As shown in Fig.~\ref{fig:temperature-dir}, the orange surface represents $T_{\text{dir}}$ as a function of $\alpha$ and $\beta$, while the blue plane indicates a ceiling temperature of the order of $10^{-8}$ K. If $(\alpha,\beta)$ are such that $T_{\text{dir}}$ lies above this ceiling, then the Bose-Einstein condensate experiment in \cite{Deppner:2021fks} should have exhibited fractional quantum mechanical behavior, since $10^{-12}$ K $\ll T_{\text{dir}}$. Thus, we can infer that the region where $T_{\text{dir}}$ peaks above the ceiling is experimentally excluded.

We can further analyze the two-dimensional parameter space $(\alpha,\beta)$ that lies within the excluded region (i.e., the peaked region in Fig.~\ref{fig:temperature-dir}) for different reference ceiling temperatures, as shown in Fig.~\ref{fig:excluded}. For instance, if we set $T_{\text{dir}} = 10^{-8}$ K, we can conservatively claim that the region within the blue curve is experimentally excluded. Specifically, for corrections that scale as $\sqrt{\ell_{\text{P}}}$ $(\beta=0.5$), the interval $\alpha \in (0.6, 1.6)$ would be forbidden. As we achieve lower temperatures without observing quantum gravity effects, the exclusion region expands, further constraining the allowed parameter space. This behavior is depicted in Fig.~\ref{fig:excluded} for $T_{\text{dir}} = 10^{-12}$ K and $10^{-15}$ K. Although these ceilings are optimistic, they are presented here for illustrative purposes.

\begin{figure}[h]\label{fig:excluded}
    \centering
    \includegraphics[width=\columnwidth]{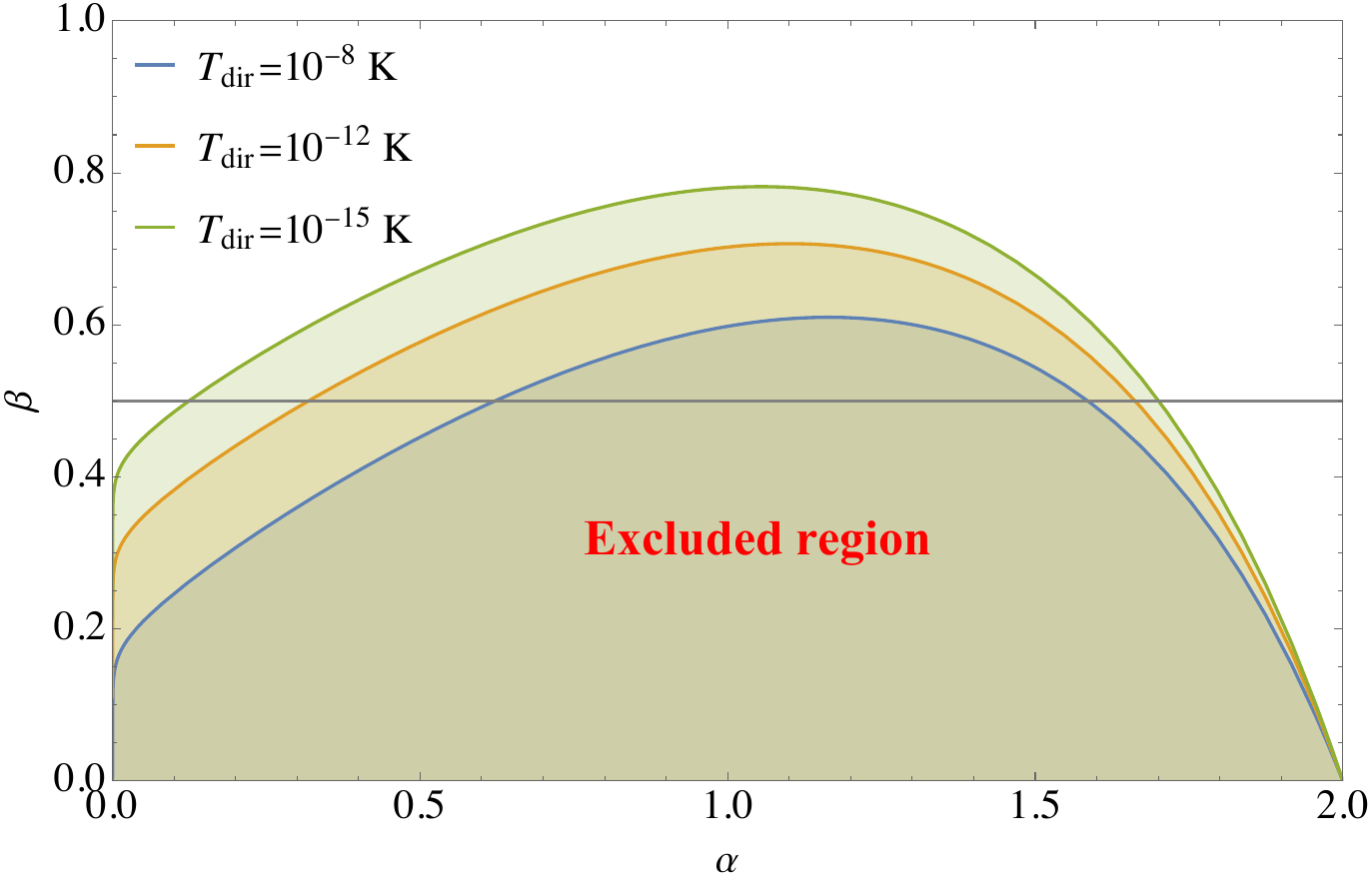}

\caption{The exclusion regions formed by the $\alpha$ and $\beta$ parameters for the temperatures $T_{\text{dir}}$. }

\label{fig:excluded}
\end{figure}

\subsection{Fractional harmonic dimension}

Regarding pure fractional quantum mechanics, several results have been derived, including the spectrum of the harmonic oscillator and the hydrogen atom \cite{Laskin:2002zz}. For the harmonic oscillator with a potential $V(x) = q^2 x^2$, the Bohr-Sommerfeld quantization rule, $2\pi \hbar \left(n + \frac{1}{2}\right) = \oint p \, dx$, can be employed to calculate the spectrum of the fractional harmonic oscillator in one spatial dimension. In this case, the energy eigenvalues are given by
\begin{equation}\label{eq:eigenene1}
    E_{n,1}^{(\alpha)} = \left( \frac{\pi \hbar D_{\alpha}^{(\beta)^{1/\alpha}} q}{B\left(\frac{1}{2}, \frac{1}{\alpha} + 1\right)} \right)^{\frac{2\alpha}{2+\alpha}} \left(n + \frac{1}{2}\right)^{\frac{2\alpha}{2+\alpha}},
\end{equation}
where the subscript $1$ indicates one dimension, and the $B$-function is defined as 
\begin{equation*}
    \frac{1}{\gamma} B\left(\frac{1}{\gamma}, \frac{1}{\alpha} + 1\right) = \int_0^1 dy \left(1 - y^{\gamma}\right)^{1/\alpha}.
\end{equation*}

Observe that Eq.\eqref{eq:eigenene1} reduces to the standard energy spectrum when $\alpha = 2$, highlighting the dimensional dependence of the oscillator:
\begin{equation}\label{eq:eigenene2}
    E_{n}^{(2)} = 2 \hbar q \sqrt{D_2^{(\beta)}} \left(n + \frac{d}{2}\right), \qquad d = 1,
\end{equation}
such that if $D_2^{(\beta)} = 1/(2M)$ and $q = \omega \sqrt{M/2}$, we obtain the textbook energy levels $E_n^{(2)} = \hbar \omega (n + d/2)$, where $d$ is the oscillator dimension, which in this case is $d=1$.




In the standard case, knowing the energy spectrum $E_n^{(2)}$ and the harmonic frequency $\omega$ allows us to determine the oscillator's dimension using the formula $d = 2E_0 / (\hbar \omega)$. However, this property is not valid in the fractional case. The concept of dimension becomes tricky in fractional quantum mechanics, where a fractal dimension often emerges naturally (see, e.g., the fractal dimension of the black hole and cosmological horizons \cite{Barrow:2020tzx,Wang:2022hun,Jalalzadeh:2021gtq,Junior:2023fwb,deOliveiraCosta:2023srx}, and the fractal dimension of Lévy paths \cite{laskin2018fractional}). Thus, a similar phenomenon might occur in this approach.


The natural candidate for an undeformed observable that depends on the system's dimension is the energy levels of the oscillator. To derive a meaningful formula for the dimension, we compare the deformed energy levels with the undeformed ones in $d$ dimensions, depending only on the Lévy index $\alpha$. Using the dimensionless formulation of fractional quantum mechanics and the energy levels of the harmonic oscillator from section 9.3 of \cite{laskin2018fractional}, we define the dimensionless energy as
\begin{align}\label{eq:dimless_ene1}
\epsilon^{(\alpha)}_{n,1} & = \frac{E_{n,1}^{(\alpha)}}{\left((D_{\alpha}^{(\beta)})^2 \hbar^2 q^{2\alpha}\right)^{1/(2+\alpha)}}\nonumber \\ 
& = \left[\frac{\pi}{B\left(\frac{1}{2}, \frac{1}{\alpha} + 1\right)} \left(n + \frac{1}{2}\right)\right]^{2\alpha/(2+\alpha)}.
\end{align}

This expression should be compared with the case $\alpha = 2$ in $d_{\alpha}$ dimensions, given by
\begin{equation}\label{eq:dimless_ene2}
    \epsilon^{(\alpha)}_{n,d_{\alpha}} = \frac{2 \hbar q \sqrt{D_2^{(\beta)}}}{\left((D_2^{(\beta)})^2 \hbar^4 q^4\right)^{1/4}} \left(n + \frac{d}{2}\right) = 2 \left(n + \frac{d_{\alpha}}{2}\right).
\end{equation}

Equating \eqref{eq:dimless_ene1} and \eqref{eq:dimless_ene2} as $\epsilon^{(\alpha)}_{n,1} = \epsilon^{(\alpha)}_{n,d_{\alpha}}$ and setting $n = 0$ for simplicity, we find
\begin{equation}
    d_{\alpha} = \left[\frac{\pi}{B\left(\frac{1}{2}, \frac{1}{\alpha} + 1\right)} \left(n + \frac{1}{2}\right)\right]^{2\alpha/(2+\alpha)}.
\end{equation}

\begin{figure}[!h]
    \centering
    \includegraphics[scale=0.67]{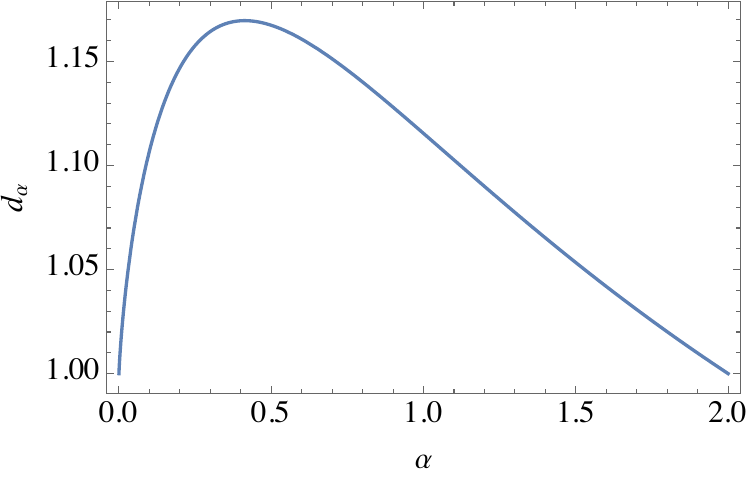}
    \caption{Harmonic fractional dimension $d_{\alpha}$ as a function of the Lévy index $\alpha$.}
    \label{fig:realdimension-dir}
\end{figure}

The dependence of this dimension on $\alpha$ is depicted in Fig.\ref{fig:realdimension-dir}. At this extreme regime, this dimension depends solely on the MDR power and not on the system's physical properties. This feature also occurs in the spectral and thermal dimensions within the UV limit \cite{Amelino-Camelia:2013tla,Amelino-Camelia:2016sru,Lobo:2020oqb}. This quantity indeed resembles a fractal dimension, as it is slightly larger than $1$ for $0 < \alpha < 2$, approaching a 2-dimensional geometry, similar to the Koch curve \cite{feder1988fractals,laskin2018fractional}. The maximum value is $d_{0.413} \approx 1.167$, and the linear case gives $d_1 \approx 1.115$.

Given that similar fluctuations occur in the UV regime, this may represent an interplay between the trans-Planckian UV and the deep IR behaviors, where similar properties emerge.

\section{Experimental analogue}\label{sec:analogue}

Despite the challenges in testing the existence of fractional quantum mechanics at the Planck scale, we can gain insights by translating experiments conducted with fractional quantum systems to the quantum gravity context. In \cite{Liu:2022xev}, an experiment was conducted using an optical apparatus in the temporal domain to realize a fractional Schrödinger equation. The Schrödinger equation considered in \cite{Liu:2022xev} is
\begin{equation}\label{eq:schr_exp1}
    i\frac{\partial}{\partial z}\Psi = \left[-\frac{|\beta_2|}{2}\frac{\partial^2}{\partial \tau^2} + i\frac{\beta_3}{6}\frac{\partial^3}{\partial \tau^3} + \frac{D}{2}\left(-\frac{\partial^2}{\partial \tau^2}\right)^{\alpha/2}\right]\Psi,
\end{equation}
where $\beta_2$ and $\beta_3$ are experimental parameters, $D$ is the fractional dispersion coefficient, $z$ is the propagation distance and $\tau=T-z/\vg$ is the reduced time, where $T$ is the physical time and $V_{\text{gr}}$ is the group velocity of the carrier wave \cite{2024Chaos..34b2102M}. This equation describes the slowly varying amplitude $\Psi$ of the electric field triggered by the travel of an optical pulse through a complex dispersive material.

In such optics experiments, the group velocity of the carrier wave $V_{\text{gr}}$ is the natural velocity scale, allowing us to map the propagation and temporal coordinates $(\tau,z)$ to new coordinates $(t,x)$ that will be used in our analysis as $z=t\vg$ and $\tau=x/\vg$ in \eqref{eq:schr_exp1}. This group velocity is found from the relation between the wavenumber $k$ and frequency $\omega$ of the carrier as $k=(D/2)|\omega|^{\alpha}-(\beta_2/2)\omega^2-(\beta_3/6)\omega^3$, which implies
\begin{equation}\label{eq_vgr1}
    \frac{1}{\vg(\alpha)}=\frac{dk}{d\omega}=\frac{\alpha D}{2}|\omega(k)|^{\alpha-1}-\beta_2|\omega(k)|-\frac{\beta_3}{2}|\omega(k)|^2\, ,
\end{equation}
where $|\omega(k)|$ is the frequency as a function of the wavenumber $k$ found from the dispersion relation. Therefore, considering fixed wavelength $\lambda=1/k$ and other experimental parameters $D$ and $\beta_{2,3}$, for each $\alpha$, we have a different $\omega$ and, consequently, a different $\vg$.

If we proceed with this map and multiply \eqref{eq:schr_exp1} by $\hbar$, it goes to
\begin{align}\label{eq:schr_exp2}
    i\hbar\frac{\partial}{\partial t}\Psi = &\left[-\frac{\hbar|\beta_2|\vg^3}{2}\frac{\partial^2}{\partial x^2} + i\frac{\hbar\beta_3\vg^4}{6}\frac{\partial^3}{\partial x^3}\right.\\\nonumber 
    &\left.+ \frac{\hbar D|\vg|^{\alpha+1}}{2}\left(-\frac{\partial^2}{\partial x^2}\right)^{\alpha/2}\right]\Psi\, .
\end{align}

This expression is equivalent to
\begin{align}
    i\hbar\frac{\partial}{\partial t}\Psi &= \left[-\frac{\hbar^2}{2M}\frac{\partial^2}{\partial x^2} - i\frac{\xi_3\hbar^3}{2Mp_{\text{P}}}\frac{\partial^3}{\partial x^3}\right.\nonumber \\ &\left.+ \xi_{\alpha,\beta}\p^{-\beta}\frac{(Mc)^{3-\alpha+\beta}\hbar^{\alpha}}{2M}\left(-\frac{\partial^2}{\partial x^2}\right)^{\alpha/2}\right]\Psi,
\end{align}
where the parameters are identified as 
\begin{align}\label{eq:map1}
    &M = \frac{\hbar}{\vg^3}\frac{1}{|\beta_2|}, \qquad \xi_3 = -\frac{M\vg^4 p_{\text{P}}}{3\hbar^3}\beta_3, \nonumber \\ &\xi_{\alpha,\beta} = \p^{\beta} M \vg^{\alpha+1}\hbar^{1-\alpha}(Mc)^{\alpha-\beta-2}D.
\end{align}

Except for the term proportional to $\xi_3$, this equation is equivalent to the one described in this letter, given by Eq.\eqref{eq:main_schr}. Hence, the experiment performed in \cite{Liu:2022xev} can be considered an analog of testing the Lorentz-violating Schrödinger equation with an optical system, representing an analogue model of in-vacuo dispersion.

In this letter, we shall focus in the case in which $\beta_3=0$. The experimental parameters $\alpha\, , \beta_2\, ,D$ and the wavelength of the carrier wave $\lambda$ are
\begin{align}\label{eq:exp_par1}
    &|\beta_2| = 21 \times 10^{-3} \, \text{p}\text{s}^2\text{/m}, \qquad D = 21 \times 10^{-3} \, \text{ps}^{\alpha}/\text{m}, \nonumber \\ 
    &\lambda= 810\, \text{nm},\qquad\alpha = 0.25, \quad \alpha = 1.25.
\end{align}

In both cases, the experimental results matched the behavior predicted by simulations. The initial pulse, represented by a Gaussian function in time, showed delocalization after traversing the nonlinear medium. For $\alpha = 1.25$, the wave packet split and displayed an interference pattern, while for $\alpha = 0.25$, a soliton rain profile emerged, with each soliton centered around different time values. This system, analogous to a wave packet traveling in quantum spacetime, suggests that a wave packet would exhibit spatial delocalization in a quantum spacetime. Such nonlocal behavior is not surprising, as the notion of locality in quantum gravity has been questioned \cite{Amelino-Camelia:2011lvm}.

The frequency of the carrier wave can be found from the dispersion relation, but due to the magnitude of the experimental parameters, it is basically the same for both $\alpha$'s, being $\omega\approx 1.1\times 10^{10}\, \text{s}^{-1}$. Using this information in \eqref{eq_vgr1}, we have $\vg=4.4\, \text{km}/{s}$. Using the experimental parameters in \eqref{eq:exp_par1} and the mapping \eqref{eq:map1}, this optical system effectively describes a particle with mass $M = 5.9 \times 10^{-32}$ kg, which is roughly of the order of magnitude of the electron mass $M_e= 9\times 10^{-31}$ kg.

The experimentally analyzed cases correspond to the dimensionless parameters $\xi_{0.25,1} =3.5 \times 10^{6}$ and $\xi_{1.25,1} =2.6 \times 10^{9}$. In order to have $\xi_{\alpha,\beta}\approx 1$ (which would be natural for quantum gravity effects), we need to have $\beta\sim 0.75$ (for $\alpha=0.25$) and $\beta\sim 0.6$ (for $\alpha=1.25$). 
Therefore, our conclusion is that such experiments are actually testing the fractal nature of spacetime, by testing in-vacuo dispersion due to quantum gravity for fractional powers in the momentum and fractional powers in the Planck scale. 

Simulations show wave packet splitting in both scenarios, and we anticipate similar experimental results. Furthermore, in the analysis carried out in \cite{Liu:2022xev}, such splitting increases with propagation, indicating an effect that accumulates over time in our analogy.

\section{Final remarks}\label{sec:final}

We revisited the modified Schrödinger equation due to modified dispersion relations. Generalizing this equation to include corrections with noninteger powers in momentum, we found it to be well suited to the realm of fractional quantum mechanics, a rich area of research with numerous experimental applications.

This identification prompted us to conjecture the existence of a deep infrared regime of quantum gravity, contrasting with the transplanckian regime in the UV. Analogous to high-energy probes, the corrected terms become more relevant than the Galilean relativistic one, leading to the emergence of fractal structures akin to those observed in the UV. In this regime, the dimension of the physical system depends on the specifics of the theory under scrutiny. We also briefly discussed the temperature scale at which such a regime would be relevant, presenting technological challenges for scrutinizing it with Planck scale sensitivity. In particular, we conjectured how to derive exclusion regions for the parameters considered when analyzing quantum systems at very low temperatures.

Also, by identifying a Planck-scale modified Schrödinger equation with fractional quantum mechanics, we recognized that experiments testing this approach serve as analog models of quantum gravity, where nonlocal behavior of the wave function emerges. 

\acknowledgments
I. P. L. was partially supported by the National Council for Scientific and Technological Development - CNPq grant 312547/2023-4 and by the grant 3197/2021, Para\'iba State Research Foundation (FAPESQ). V. B. B. was partially supported by the National Council for Scientific and Technological Development - CNPq grant 307211/2020-7. G .V. was supported by the National Council for Scientific and Technological Development - CNPq grant 140335/2022-6.

\bibliographystyle{eplbib}       
\bibliography{dir}   

\begin{thebibliography}{10}
\expandafter\ifx\csname url\endcsname\relax\def\url#1{\texttt{#1}}\fi

\bibitem{Kiefer:2004xyv}
\Name{Kiefer C.} \Book{{Quantum gravity}} Vol. 124 (Clarendon, Oxford) 2004.

\bibitem{Polchinski:1998rq}
\Name{Polchinski J.} \Book{{String theory. Vol. 1: An introduction to the
  bosonic string}} Cambridge Monographs on Mathematical Physics (Cambridge
  University Press) 2007.

\bibitem{Ashtekar:2021kfp}
\Name{Ashtekar A. \and Bianchi E.} \REVIEW{Rept. Prog.
  Phys.}{84}{2021}{042001}.

\bibitem{Loll:2019rdj}
\Name{Loll R.} \REVIEW{Class. Quant. Grav.}{37}{2020}{013002}.

\bibitem{Eichhorn:2018yfc}
\Name{Eichhorn A.} \REVIEW{Front. Astron. Space Sci.}{5}{2019}{47}.

\bibitem{Amelino-Camelia:2008aez}
\Name{Amelino-Camelia G.} \REVIEW{Living Rev. Rel.}{16}{2013}{5}.

\bibitem{Addazi:2021xuf}
\Name{Addazi A. \etal} \REVIEW{Prog. Part. Nucl. Phys.}{125}{2022}{103948}.

\bibitem{Mattingly:2005re}
\Name{Mattingly D.} \REVIEW{Living Rev. Rel.}{8}{2005}{5}.

\bibitem{AlvesBatista:2023wqm}
\Name{Alves~Batista R. \etal} \REVIEW{arXiv:2312.00409}{}{2023}{}.

\bibitem{Wagner:2023fmb}
\Name{Wagner F., Var\~ao G., Lobo I.~P. \and Bezerra V.~B.} \REVIEW{Phys. Rev.
  D}{108}{2023}{066008}.

\bibitem{Sorkin:1996sr}
\Name{Sorkin R.~D.} \Book{{How wrinkled is the surface of a black hole?}}
  presented at \Book{{1st Australasian Conference (ACGRG1) on General
  Relativity and Gravitation (Gravitational Waves, Mathematical Relativity,
  Quantum Gravity)}} 1996.

\bibitem{Marolf:2003bb}
\Name{Marolf D.} \REVIEW{Springer Proc. Phys.}{98}{2005}{99}.

\bibitem{Keshet:2024hlm}
\Name{Keshet E., Shemesh I. \and Steinhauer J.}
  \REVIEW{arXiv:2407.00448}{}{2024}{}.

\bibitem{Yunes:2016jcc}
\Name{Yunes N., Yagi K. \and Pretorius F.} \REVIEW{Phys. Rev.
  D}{94}{2016}{084002}.

\bibitem{Laskin:1999tf}
\Name{Laskin N.} \REVIEW{Phys. Lett. A}{268}{2000}{298}.

\bibitem{laskin2000fractional}
\Name{Laskin N.} \REVIEW{Physical Review E}{62}{2000}{3135}.

\bibitem{laskin2018fractional}
\Name{Laskin N.} \Book{Fractional Quantum Mechanics} (WORLD SCIENTIFIC) 2018.

\bibitem{Lukierski:1992dt}
\Name{Lukierski J., Nowicki A. \and Ruegg H.} \REVIEW{Phys. Lett.
  B}{293}{1992}{344}.

\bibitem{Majid:1994cy}
\Name{Majid S. \and Ruegg H.} \REVIEW{Phys. Lett. B}{334}{1994}{348}.

\bibitem{Amelino-Camelia:2016gfx}
\Name{Amelino-Camelia G., da~Silva M.~M., Ronco M., Cesarini L. \and Lecian
  O.~M.} \REVIEW{Phys. Rev. D}{95}{2017}{024028}.

\bibitem{Amelino-Camelia:1996bln}
\Name{Amelino-Camelia G., Ellis J.~R., Mavromatos N.~E. \and Nanopoulos D.~V.}
  \REVIEW{Int. J. Mod. Phys. A}{12}{1997}{607}.

\bibitem{Freidel:2005me}
\Name{Freidel L. \and Livine E.~R.} \REVIEW{Phys. Rev.
  Lett.}{96}{2006}{221301}.

\bibitem{Amelino-Camelia:2009wvc}
\Name{Amelino-Camelia G., Laemmerzahl C., Mercati F. \and Tino G.~M.}
  \REVIEW{Phys. Rev. Lett.}{103}{2009}{171302}.

\bibitem{Barrow:2020tzx}
\Name{Barrow J.~D.} \REVIEW{Phys. Lett. B}{808}{2020}{135643}.

\bibitem{Adler:2001vs}
\Name{Adler R.~J., Chen P. \and Santiago D.~I.} \REVIEW{Gen. Rel.
  Grav.}{33}{2001}{2101}.

\bibitem{Amelino-Camelia:2005zpp}
\Name{Amelino-Camelia G., Arzano M., Ling Y. \and Mandanici G.} \REVIEW{Class.
  Quant. Grav.}{23}{2006}{2585}.

\bibitem{Ling:2005bq}
\Name{Ling Y., Hu B. \and Li X.} \REVIEW{Phys. Rev. D}{73}{2006}{087702}.

\bibitem{Calcagni:2009kc}
\Name{Calcagni G.} \REVIEW{Phys. Rev. Lett.}{104}{2010}{251301}.

\bibitem{Amelino-Camelia:2013tla}
\Name{Amelino-Camelia G., Arzano M., Gubitosi G. \and Magueijo J.}
  \REVIEW{Phys. Rev. D}{87}{2013}{123532}.

\bibitem{Benedetti:2008gu}
\Name{Benedetti D.} \REVIEW{Phys. Rev. Lett.}{102}{2009}{111303}.

\bibitem{Horava:2009if}
\Name{Horava P.} \REVIEW{Phys. Rev. Lett.}{102}{2009}{161301}.

\bibitem{Calcagni:2013vsa}
\Name{Calcagni G., Eichhorn A. \and Saueressig F.} \REVIEW{Phys. Rev.
  D}{87}{2013}{124028}.

\bibitem{Sotiriou:2011aa}
\Name{Sotiriou T.~P., Visser M. \and Weinfurtner S.} \REVIEW{Phys. Rev.
  D}{84}{2011}{104018}.

\bibitem{Amelino-Camelia:2016sru}
\Name{Amelino-Camelia G., Brighenti F., Gubitosi G. \and Santos G.}
  \REVIEW{Phys. Lett. B}{767}{2017}{48}.

\bibitem{Lobo:2020oqb}
\Name{Lobo I.~P. \and Santos G.~B.} \REVIEW{Phys. Lett. B}{817}{2021}{136272}.

\bibitem{Bosso:2023nst}
\Name{Bosso P., Fabiano G., Frattulillo D. \and Wagner F.} \REVIEW{Phys. Rev.
  D}{109}{2024}{046016}.

\bibitem{Amelino-Camelia:2013fxa}
\Name{Amelino-Camelia G.} \REVIEW{Phys. Rev. Lett.}{111}{2013}{101301}.

\bibitem{Deppner:2021fks}
\Name{Deppner C. \etal} \REVIEW{Phys. Rev. Lett.}{127}{2021}{100401}.

\bibitem{Laskin:2002zz}
\Name{Laskin N.} \REVIEW{Phys. Rev. E}{66}{2002}{056108}.

\bibitem{Wang:2022hun}
\Name{Wang L.-H. \and Ma M.-S.} \REVIEW{Phys. Lett. B}{831}{2022}{137181}.

\bibitem{Jalalzadeh:2021gtq}
\Name{Jalalzadeh S., da~Silva F.~R. \and Moniz P.~V.} \REVIEW{Eur. Phys. J.
  C}{81}{2021}{632}.

\bibitem{Junior:2023fwb}
\Name{Junior P. F. d.~S., Costa E. W. d.~O. \and Jalalzadeh S.} \REVIEW{Eur.
  Phys. J. Plus}{138}{2023}{862}.

\bibitem{deOliveiraCosta:2023srx}
\Name{de~Oliveira~Costa E.~W., Jalalzadeh R., da~Silva, Junior. P.~F., Rasouli
  S. M.~M. \and Jalalzadeh S.} \REVIEW{Fractal Fract.}{7}{2023}{854}.

\bibitem{feder1988fractals}
\Name{Feder J.} \Book{Fractals} Physics of solids and liquids (Plenum Press)
  1988.

\bibitem{Liu:2022xev}
\Name{Liu S., Zhang Y., Malomed B.~A. \and Karimi E.} \REVIEW{Nature
  Commun.}{14}{2023}{222}.

\bibitem{2024Chaos..34b2102M}
\Name{{Malomed} B.~A.} \REVIEW{Chaos}{34}{2024}{022102}.

\bibitem{Amelino-Camelia:2011lvm}
\Name{Amelino-Camelia G., Freidel L., Kowalski-Glikman J. \and Smolin L.}
  \REVIEW{Phys. Rev. D}{84}{2011}{084010}.

\end{thebibliography}

%
%

\end{document}